\documentclass[aip,apl,reprint,amsmath,amssymb]{revtex4-1}

\usepackage{graphicx}
\usepackage{textcomp} 
\usepackage{epstopdf}

\begin{document}


\title{Wideband superconducting nanotube electrometer} 




\author{Pasi H\"akkinen}
\affiliation{Low Temperature Laboratory, Department of Applied Physics, Aalto University, FI-00076 AALTO, FINLAND}
\author{Aur\'elien Fay}
\affiliation{Low Temperature Laboratory, Department of Applied Physics, Aalto University, FI-00076 AALTO, FINLAND}
\author{Dmitry Golubev}
\affiliation{Low Temperature Laboratory, Department of Applied Physics, Aalto University, FI-00076 AALTO, FINLAND}
\author{Pasi L\"ahteenm\"aki}
\affiliation{Low Temperature Laboratory, Department of Applied Physics, Aalto University, FI-00076 AALTO, FINLAND}
\author{Pertti Hakonen}
\email[]{pertti.hakonen@aalto.fi}
\affiliation{Low Temperature Laboratory, Department of Applied Physics, Aalto University, FI-00076 AALTO, FINLAND}


\date{\today}

\begin{abstract}
We have investigated the microwave response of nanotube Josephson junctions at 600-900 MHz at microwave powers corresponding to currents from 0 to \(2\times I_{\mathrm C}\) in the junction. Compared with theoretical modeling, the response of the junctions correspond well to the lumped element model of resistively and capacitively shunted junction. We demonstrate the operation of these superconducting FETs as charge detectors at high frequencies without any matching circuits. Gate-voltage-induced charge \(Q_{\rm G}\) modifies the critical current \(I_{\rm C}\), which changes the effective impedance of the junction under microwave irradiation. This change, dependent on the transfer characteristics \(dI_{\mathrm C}/dQ_{\rm G}\), modifies the reflected signal and it can be used for wide band electrometry. We measure a sensitivity of \(3.1\times10^{-5}\) \(e/\sqrt{\mathrm{Hz}}\) from a sample which has a maximum switching current of 2.6 nA.
\end{abstract}


\maketitle 


Superconducting correlations can be induced across superconductor-normal (SN) metal boundaries~\cite{Gennes1964,Tinkham} and they lead to well defined Josephson effect~\cite{Josephson1962} in short, good quality SNS samples. The consequent proximity-induced supercurrents have been investigated in single walled carbon nanotubes~\cite{KasumovScience,Morpurgo,Kasumov,Herrero,Jorgenssen,Wernsdorfer,Cleuziou2007,Eichler2007,Zhang2008,Fan,Rasmussen2009,Schneider2012} as well as multiwalled tubes~\cite{Vecino,HaruyamaI,HaruyamaII,Tsuneta,Tsuneta2009,Lechner2011}, extending to
their inductive gate-modulation (Josephson inductance) in microwave cavities\cite{Lechner2011}. Superconducting carbon nanotube devices provide mesoscopic components that are at the same time moderate-impedance and charge-sensitive. This is exceptional because typically resistance of a nanosample has to be around the quantum resistance \(R_{\mathrm Q}=h/e^2\) in order to obtain charge quantization effects. The low impedance nature of such devices makes them very attractive for high frequency electrometry as the matching circuits between the samples and the 50 \(\Omega\) measuring setup can be avoided.

Here we demonstrate direct coupling of a superconducting multiwalled carbon nanotube junction to transmission line. We operate the device as a charge detector, and find a sensitivity of \(\delta Q=3.1\times10^{-5}\) \(e/\sqrt{\mathrm{Hz}}\) from direct reflection measurement. The junction, having a maximum switching current of \(I_{\mathrm  sw}=2.6\) nA, gives the best sensitivity when it is current biased close to the switching point, where it has finite source-drain voltage due to phase diffusion. The electrical properties and the operation principle as charge detector are modeled using the phenomenological resistively and capacitively shunted junction (RCSJ) model. Since the device is probed at frequency range that is well below Josephson frequency, the junction can be modeled as a resistive load having its response determined by the DC \textit{IV}-characteristics.

\begin{figure} 
\includegraphics[]{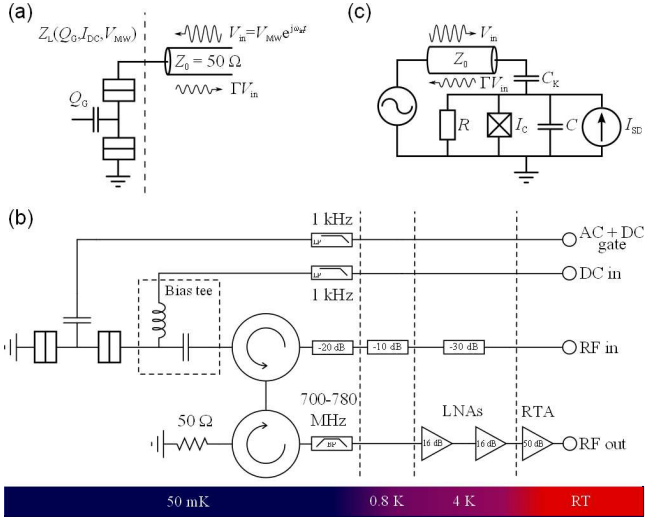}
\caption{(a) An electromagnetic wave from a transmission line is reflected from the superconducting nanotube sample; \(Z_{\mathrm L}\) entering Eq.~(\ref{eq_gamma}) is the impedance seen to the left from the dashed line. (b) Lumped element RCSJ-model of a Josephson junction that is capacitively coupled to a transmission line having characteristic impedance \(Z_{0}\). (c) Experimental realization of (a). The reflection measurement is implemented using two circulators in order to block the preamplifier (LNA) noise from reaching the sample.}
\label{fig_measurement_scheme}
\end{figure}

The basic concept of direct reflection measurement is illustrated in Fig.~\ref{fig_measurement_scheme}a. The nanotube device is connected directly to the end of a \(Z_{0}=50\) \(\Omega\) transmission line. The impedance \(Z_{\mathrm L}\) that a superconducting nanotube will display will depend strongly on the bias conditions imposed on the sample, i.e., on the gate charge \(Q_{\rm G}\), controlling Josephson energy \(E_{\rm J}\), and on the bias, both DC current \(I_{\rm DC}\) and microwave carrier voltage \(V_{\mathrm MW}\), which control the potential barrier for phase diffusion. At low levels of the carrier and bias current, the superconducting sample will look like a purely inductive load and reflect fully the incoming microwaves as controlled by the reflection coefficient
\begin{equation}
\Gamma =\frac{Z_{\mathrm L}\left(Q_{\mathrm G},I_{\rm DC},V_{\mathrm MW}\right)-Z_{0}}{Z_{\mathrm L}\left(Q_{\mathrm G},I_{\rm DC},V_{\mathrm MW}\right)+Z_{0}}. 
\label{eq_gamma}
\end{equation}
With increasing carrier amplitude and bias current, \(Z_{\mathrm L}\) will grow due to increased phase diffusion. Consequently, one can minimize the reflection by tuning the gate charge and bias, and accomplish a good matching in this way without the use of any extra matching elements.

In the radio-frequency single-electron transistor (rf-SET), for comparison, the carrier wave is reflected from an impedance transformer \textit{LC}-circuit and the high impedance SET~\cite{Schoelkopf1998}. The variation of the island charge changes the impedance of the SET, and the amplitude of the reflected wave is modulated according to these changes. The rf-SET bandwidth is limited by the loaded \textit{Q}-factor of the impedance transformer. In our superconducting nanotube scheme, the bandwidth is limited due to noise thermalization requirements and due to preamplifier band width restriction for low noise.

When the modulation of \(Z_{\rm L}\) is the dominant factor in varying \(\Gamma\), the equation for charge resolution for the superconducting nanotube read-out as well as for the regular rf-SET read-out is~\cite{Roschier2004a}
\begin{equation}
\delta Q_{\rm{RMS}}=\frac{\sqrt{2k_{\rm B}T_{\rm N}Z_{0}\nu_{\rm{bw}}}}{\frac{V_{\rm MW}}{\sqrt{2}}\frac{\partial\left|\Gamma \right|}{\partial Q_{\rm G}}},
\label{eq_sensitivity_theoretical}
\end{equation}
where it is assumed that the noise level in the system is determined by the cooled preamplifier. Here \(T_{\rm N}\) is the noise temperature of the preamplifier and \(\nu_{\rm{bw}}\) denotes the resolution bandwidth of the spectrum analyzer. For the optimum sensitivity we want to have a large carrier amplitude, large transfer function  \(\partial\left|\Gamma\right|/\partial Q_{\rm G}\), small noise temperature \(T_{\mathrm N}\), and a narrow resolution bandwidth \(\nu_{\mathrm{bw}}\). Using impedance matching circuits, MWNT rf-SETs have yielded charge sensitivities of \(2\times10^{-5}\) \(e/\sqrt{\mathrm{Hz}}\)~\cite{Roschier2004b}.

The schematics of the measurement circuitry used for the direct reflection measurement is illustrated in Fig.~\ref{fig_measurement_scheme}b. Experiment was carried out in a cryogen-free dilution refrigerator having strongly attenuated measurement leads for eliminating electronic noise from contaminating the sample. Filtering was necessary because for small Josephson energies \(\left(E_{\mathrm J}\sim200\ \mathrm{mK}\right)\) excess noise has been found to strongly suppress the measured critical current from the real \(I_{\rm C}\)~\cite{Herrero}. Charge sensitivity was measured by feeding a 700 MHz carrier signal to the source of the sample and modulating the gate at 470 Hz. The reflected signal was then taken via two circulators to two cascaded 16 dB low noise amplifiers located  at 4.2 K~\cite{amplifier}. After further amplification by 50 dB at room temperature, the output frequency spectrum was read with a spectrum analyzer and the sensitivity was directly deduced from the sideband height. 

The sample employed in these studies was fabricated in a rather standard fashion. The tube material was grown using plasma enhanced growth without any metal catalyst by the group of Sumio Iijima. The tubes were dispersed in dichloroethane and deposited onto thermally oxidized, strongly doped Si/SiO\(_2\) wafers. Tubes were located with respect to alignment markers using a scanning electron microscope. Subsequently, Ti contacts of width 400 nm were made using e-beam overlay lithography: 10 nm titanium layer in contact with the tube was covered by 70 nm Al in order to facilitate proximity induced superconductivity in Ti at subkelvin temperatures. The length of the tube section between the contacts was 250 nm. The electrically conducting body of the silicon substrate was employed as a back gate, separated from the sample by 273 nm of SiO\(_2\).

\begin{figure}
  \begin{center}
  \includegraphics[]{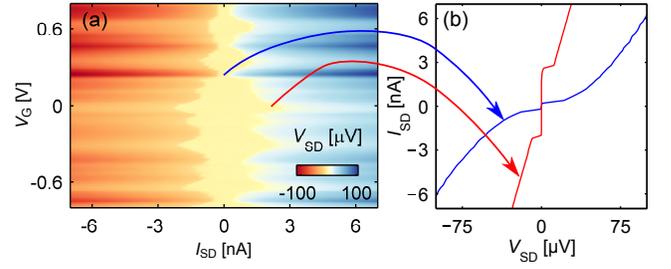}
    \caption{(a) Collection of \textit{IV}-curves measured within \(\pm 0.8\) V gate voltage range from a current biased carbon nanotube sample with Ti/Al contacts. Supercurrent is seen at all \(V_{\rm G}\) values. (b) \textit{IV}-curves measured at \(V_{\rm G}=0.24\) V, blue, and \(V_{\rm G}=-0.015\) V, red, demonstrate gate tunability of the switching and retrapping current.}
     \label{fig_dc_characteristics}
 \end{center}
\end{figure}

Figs.~\ref{fig_dc_characteristics}a and b display \textit{IV}-characteristics of our sample for different gate voltages measured using current bias. The \textit{IV}-curves display hysteresis and gate voltage tunable switching and retrapping current. The largest measured switching current was \(I_{\rm sw}=2.6\) nA and the corresponding measured retrapping current \(I_{\rm{r}}=2\) nA (red curve in Fig.~\ref{fig_dc_characteristics}b). The supercurrent was observed at all gate voltage values and it varied in the range \(I_{\rm sw}=0.2-2.6\) nA. Hence Coulomb blockade and the resulting variation of normal state resistance vs. \(V_{\rm G}\) was weak in this sample. The back gate capacitance \(C_{\rm G}=1.2\) aF can be deduced from the measured gate period of SET oscillations \(\Delta V_{\rm G}^{e}=\) 130 mV. This is a conservative estimate that was used for determining the charge sensitivity.

In order to check how large carrier powers \(P_{\rm MW}\) can be employed in the reflection measurement, we have recorded DC \textit{IV}-characteristics under different excitations. At low levels, \(P_{\rm MW}<-120\) dBm, no change in the \textit{IV}-curve is observed. First above -118 dBm, the zero-bias slope of the \textit{IV}-curve starts to become more resistive. When the current amplitude of the microwave and the influence of current noise increase the effective excitation up to \(I_{\rm C}\), a plateu is developed around the zero current bias.

\begin{figure}
  \begin{center}
  \includegraphics[]{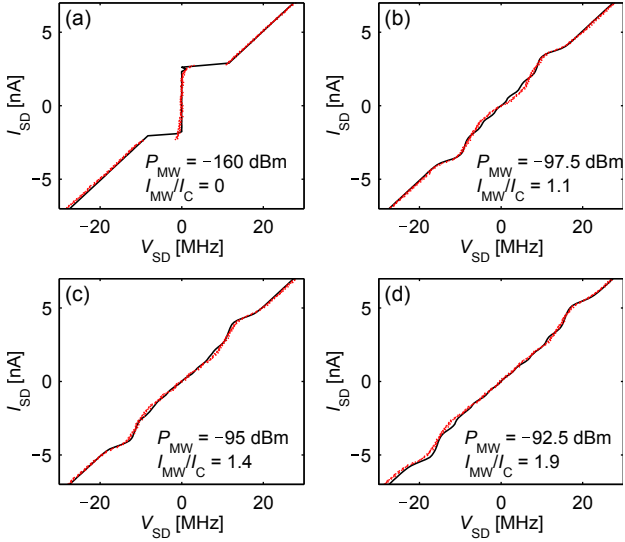}
	\caption{\(IV\) characteristics of the sample under strong microwave irradiation. Measured and simulated data are plotted with red and black color, respectively. Sample is biased to \(V_{\rm G}=-0.015\) V. Equivalent microwave current amplitudes \(I_{\rm MW}/I_{\rm C}\) are 0, 1.1, 1.4, and 1.9 in (a), (b), (c), and (d), respectively. The junction parameters were chosen as: \(C+C_{\rm K}=125\) fF, \(R=3.9\) k\(\Omega\), and \(I_{\rm C}=5.7\) nA. The total junction capacitance is thus dominated by the coupling capacitor \(C_{\rm{K}}\).}
     \label{fig_iv_vs_power}
 \end{center}
\end{figure}

The \textit{IV}-characteristics plotted with red in Fig.~\ref{fig_iv_vs_power} have been measured at \(V_{\rm G}=-0.015\) V. The switching current was maximum at this bias point in the absence of microwave excitation. Equation describing the response of a current biased Josephson junction that is capacitively coupled to a transmission line through \(C_{\rm K}\), see Fig.~\ref{fig_measurement_scheme}(c), takes a form~\cite{Tinkham}
\begin{equation}
(C+C_{\rm K})\frac{\hbar\ddot\varphi}{2e}+\frac{1}{R}\frac{\hbar\dot\varphi}{2e} + I_{\rm C}\sin\varphi  = I_{\rm DC} + 2C_{\rm K} \dot V_{\rm in}(t).
\label{RSJ2}
\end{equation}
Here \(C\) and \(R\) are the junction capacitance and resistance, respectively, \(\varphi\) the phase difference over the junction, \(I_{\rm DC}\) the DC current bias and \(V_{\rm in}(t)\) the rf-excitation applied to the source electrode. Assuming that the incoming rf-excitation is sinusoidal \(V_{\rm in}(t)= V_{\rm MW}\sin\omega_{\rm in}(t)\), we have \(2C_{\rm K} \dot V_{\rm in}(t)=I_{\rm MW}\cos{\omega_{\rm in}t}\), where \(I_{\rm MW}=2\omega_{\rm in}C_{\rm K}V_{\rm{MW}}\) is the effective amplitude of current oscillations. \textit{IV}-curves were obtained by numerically solving Eq. (\ref{RSJ2}) with Gaussian white current noise, having standard deviation of \(0.125\times I_{\rm C}\), applied to each time step of the simulation. Resulting curves are presented with black color in Fig.~\ref{fig_iv_vs_power}. The measured curves fit well with the simple, approximative RCSJ-model, especially at high drives where the phase diffusion due to noise does not any more play so significant role on determining the transport properties~\cite{Vion1996,Cleuziou2007}.


The frequency of incoming radiation \(\omega_{\rm in}=700\) MHz is rather low. It becomes lower than  the Josephson frequency \(\omega_{\rm J}=2eV_{\rm SD}/\hbar\) at source-drain voltages higher than 0.2 \textmu V. Because we have \(V_{\rm SD}> 1\) \textmu V at the most sensitive charge detection points, one can average the Eq.~(\ref{RSJ2}) over time period which is longer than \(1/\omega_{\rm J}\) but shorter than \(1/\omega_{\rm in}\). Then Eq.~(\ref{RSJ2}) acquires a simple form
\begin{equation}
(C+C_K)\frac{\hbar\ddot\varphi}{2e}+I_{\rm SD}\left(V_{\rm SD},I_{\rm C}\right)  = I_{\rm DC} + 2C_{\rm K} \dot V_{\rm in}(t),
\end{equation}
where \(I_{\rm SD}(V_{\rm SD},I_{\rm C})\) is the DC \textit{IV}-curve of the junction. It is a function of two parameters: voltage \(V_{\rm SD}\) and critical current \(I_{\rm C}\). We assume that the amplitude of the incoming radiation \(I_{\rm MW}\) is small and can be treated perturbatively. In addition, we allow the modulation of the critical current, \(I_{\rm C}(t)=I_{\rm C}+\delta I_{\rm C}\cos\Omega t\), where \(\Omega=470\) Hz is a very low modulation frequency. Then
\(
\hbar\dot\varphi/2e = V_{\rm SD} + \hbar\delta\dot\varphi/2e,
\)
where $\delta\varphi$ is the small correction to the phase. In the linear response regime one finds
\begin{equation}
(C+C_{\rm K})\frac{\hbar\delta\ddot\varphi}{2e} 
+ \frac{\partial I_{\rm SD}(V_{\rm SD},I_{\rm C}(t))}{\partial V_{\rm SD}}\frac{\hbar\delta\dot\varphi}{2e} 
= 2C_{\rm K} \dot V_{\rm in}(t).
\end{equation}
In this approximation the effective impedance of the nanotube, 
$Z_L=\left[1/\left(-i\omega C_{\rm K}\right)+1/\left(-i\omega C+{\partial I_{\rm SD}}/{\partial V_{\rm SD}}\right)\right]^{-1}$,
slowly changes in time. 
As a result, the reflected wave acquires the form
\(V_{\rm out}(t) = \left( \Gamma + \Gamma_1 \cos\Omega t \right) V_{\rm in}e^{-i\omega_{\rm in}t}\), where 
$\Gamma_1$ is the reflection coefficient to the sideband. At small $\delta I_{\rm C}$  it reads
\begin{equation}
\Gamma_1 = -\frac{Z_0Z_{\rm L}^2}{(Z_{\rm L}+Z_0)^2}\frac{\partial^2 I_{\rm SD}}{\partial V_{\rm SD}\partial I_{\rm C}}\delta I_{\rm C}. 
\label{eq_sideband}
\end{equation}

\begin{figure}
  \begin{center}
  \includegraphics[]{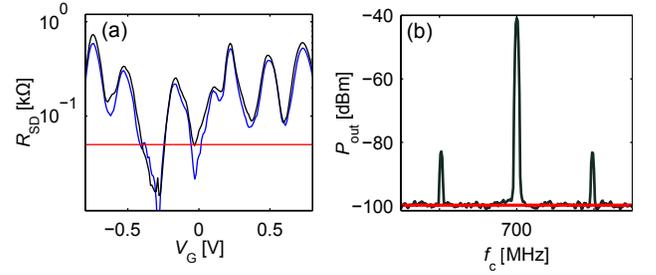}
	\caption{(a) Differential conductance measured with zero bias against \(V_{\rm G}\) with and without microwave are plotted with black and blue curves, respectively. Note that the resistance of the sample can be still tuned to 50 \(\Omega\) (red line) with the used microwave excitation of \(-120\) dBm. (b) Reflection measurement of the charge sensitivity at gate voltage point where \(I_{\rm sw}\approx1.3\) nA yields sensitivity of \(3.1\times10^{-5}\ e/\sqrt{\mathrm{Hz}}\). Carrier power of \(-120\) dBm was used for this measurement.
}
	\label{fig_sidebands}
	\end{center}
\end{figure}

Fig.~\ref{fig_sidebands}a illustrates the change in the modulation of the zero-bias resistance when the Josephson junction is exposed to microwave irradiation. Note that \(R_{\mathrm{SD}}\) still reaches 50 \(\Omega\) at regions where \(I_{\rm sw}\approx2\) nA with the applied power level of -120 dBm. Conductance characteristics do not degrade significantly at this power level and there is still a very clear modulation in \(R_{\mathrm{SD}}\) as a function of \(V_{\rm G}\).


Sideband measurement scheme was used for directly determining the charge sensitivity of the sample. Because obtaining a large signal at sideband frequency requires nonzero \(\partial^2 I_{\rm SD}/\left(\partial V_{\rm SD}\partial I_{\rm C}\right)\), the best sensitivities were found slightly off from the charge degeneracy points which have the largest switching currents. Fig.~\ref{fig_sidebands}b shows a response curve measured using a 700 MHz carrier and 470 Hz gate modulation at a gate bias corresponding to \(I_{sw}\approx1.3\) nA. We find a charge sensitivity of \(3.1\times10^{-5}\) \(e/\sqrt{\mathrm{Hz}}\) at \(P_{\rm {MW}}=-120\) dBm at the sample.

The best sensitivity was reached by biasing the sample just below the switching point, and tuning the gate for maximum signal while still staying in the superconducting state. This is in line with the relatively small value of \(C_{\rm{K}}\) needed for reproducing \textit{IV}-curves. Having a small capacitor in series with the Josephson junction causes the most sensitive operation point to be shifted from 50 \(\Omega\) to higher resistance values located close to the switching point. The sensitivity estimate of the device obtained from Eq.~\ref{eq_sideband} using measured \textit{IV}-curves reproduces the bias point location of the best operation and it also gives relatively good agreement with the measured sideband amplitude.

In conclusion, we have demonstrated that using a superconducting nanotube sample with switching current of around 2.6 nA, it is possible to construct an rf-electrometer that provides charge sensitivity of  \(3.1\times10^{-5}\) \(e/\sqrt{\mathrm{Hz}}\) at carrier frequency of 700 MHz without any matching circuitry. The  operation bandwidth is limited by the performance of the circulators and of the preamplifier.


%
%

%

\begin{acknowledgments}
We thank T. Heikkil\"a and E. Sonin for fruitful discussions. Our work was supported by the Academy of Finland (contract no. 250280, LTQ CoE) and the EU CARDEQ contract FP6-IST-021285-2 and we benefitted from the use of the Aalto University infrastructures Low Temperature Laboratory and Nanomicroscopy Center.
\end{acknowledgments}

\bibliography{references}

\end{document}